\documentclass[%
 reprint,
 amsmath,amssymb,
 aps,
longbibliography
]{revtex4-1}
\usepackage{graphicx}
\usepackage{dcolumn}
\usepackage{bm}
\usepackage{multirow}

\usepackage{amsmath,amssymb,amsfonts}
\usepackage{algorithmic}
\usepackage{graphicx}
\usepackage{textcomp}
\usepackage{xcolor}
\usepackage{color,soul}
\usepackage{nccmath}
\usepackage{braket}
\usepackage{physics}
\usepackage{caption}
\usepackage{subcaption}
\usepackage{chngcntr}

\begin{document}

\preprint{APS/123-QED}

\title{Quantum Keyless Privacy vs. Quantum Key Distribution for Space Links}

\author{A. V\'azquez-Castro$^1$ }
\author{D. Rusca$^2$ }
\author{H. Zbinden$^2$ }

\affiliation{$^1$Autonomous University of Barcelona and Center of Space Studies and Research, CERES (IEEC-UAB). Campus de Bellaterra, 08290 Barcelona, Spain\\
$^2$Group of Applied Physics, Univ. of Geneva, Chemin de Pinchat 22, CH-1211 Geneva 4, Switzerland
}%




\date{\today}

\begin{abstract}

We study information theoretical security for space links between a satellite and a ground-station. Quantum key distribution (QKD) is a well established method for information theoretical secure communication, giving the eavesdropper unlimited access to the channel and technological resources only limited by the laws of quantum physics. But QKD for space links is extremely challenging, the achieved key rates are extremely low, and day-time operating impossible. However, eavesdropping on a channel in free-space without being noticed seems complicated, given the constraints imposed by orbital mechanics. If we also exclude eavesdropper's presence in a given area around the emitter and receiver, we can guarantee that he has only access to a fraction of the optical signal. In this setting, quantum keyless private (direct) communication based on the wiretap channel model is a valid alternative to provide information theoretical security. Like for QKD, we assume the legitimate users to be limited by state-of-the-art technology, while the potential eavesdropper is only limited by physical laws: physical measurement (Helstrom detector) and quantum electrodynamics (Holevo bound). Nevertheless, we demonstrate information theoretical secure communication rates (positive keyless private capacity) over a classical-quantum wiretap channel using on-off-keying of coherent states. We present numerical results for a setting equivalent to the recent experiments with the Micius satellite and compare them to the fundamental limit for the secret key rate of QKD. We obtain much higher rates compared with QKD with exclusion area of less than 13 meters for Low Earth Orbit (LEO) satellites. Moreover, we show that the wiretap channel quantum keyless privacy is much less sensitive to noise and signal dynamics and daytime operation is possible.

\end{abstract}

\maketitle


\section{Introduction}

\begin{figure}[tbh]
\centering
\includegraphics[scale=0.28]{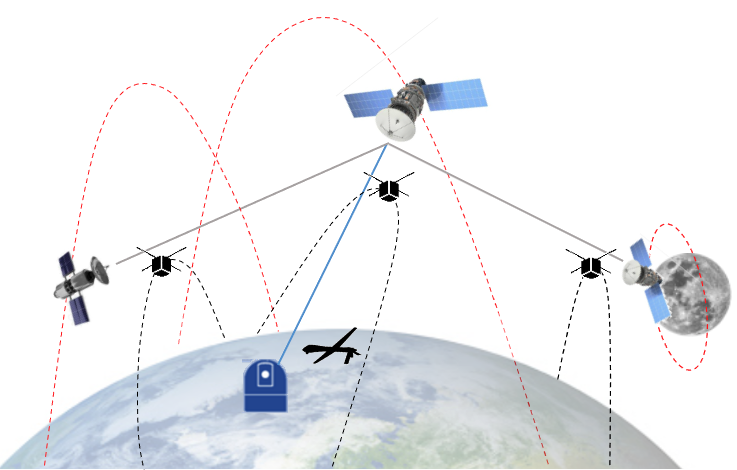}
\protect\caption{Our scenarios of interest: celestial mechanics dictates orbit dynamics for legitimate parties and eavesdroppers.} 
\label{fig:OWsecrecy_QT}
\end{figure}

The first protocol of Quantum Key Distribution (QKD) was proposed in 1984 by Bennet and Brassard \cite{BB1984}. Today, commercial fiber-based systems are available with operational distances steadily increasing \cite{Zbinden2018}\cite{Fang2020}. Recently, QKD has also been realized between the Chinese satellite Micius and ground-stations \cite{Micius2017}\cite{Micius2020}. Currently, there are several national initiatives to deploy a QKD network, including terrestrial and space links. In particular, there is an increasing and unprecedented interest in space telecommunication networks; our scenarios of interest as illustrated in Fig. \ref{fig:OWsecrecy_QT}, (this work focuses on the blue link). 

A QKD protocol assumes that two distant legitimate parties, Alice and Bob have at their disposal the following resources: Secure offices, access to an untrusted quantum channel and an authenticated classical public channel, and perfectly random numbers. A potential eavesdropper, Eve, has the complete control of the quantum channel. In particular, she can intercept all quantum states and perform any measurement allowed by quantum mechanics, including entangling the states with some auxiliary system and store them in perfect quantum memories. Moreover, she can send signals to Bob over a lossless channel. Despite this extreme power only limited by the laws of quantum physics, QKD protocols are now well established and furnished with security proofs \cite{Shor2000}\cite{Mayers2001}\cite{Kato2016}\cite{Curty2019}. So the generated keys guarantee together with the encoding using the one-time-pad information theoretically secure communication \cite{Renner2005}\cite{Scarani2009}.  

Alternatively, physical-layer-security relies also on information theoretical principles in the sense that Eve has unlimited computing power. However, she has a physical disadvantage with respect to the legitimate users. In particular, Eve can intercept only a fraction of the signal sent by Alice. In 1975,  Wyner first proposed a protocol based on a so-called degraded wiretap channel (for a  classical channel)\cite{Wyner1975}\cite{WynerPoisson}\cite{WynerPoisson-1}. The information leakage to the eavesdropper is determined with an information theoretical measure, which has been strengthened from Wyner's (weak) security to the so called strong security criterion by I.~Csisz\'ar and J.~K\"orner, \cite{Csiszar1978} and Maurer \cite{Maurer1993} until achieving cryptographic semantic security \cite{Bellare2012}.
In a quantum setting, where Alice and Bob use an authenticated quantum channel for (keyless) secure direct communication, the asymptotically achievable rate, the private capacity, subject to the strong security criterion, was determined by Cai, Winter and Yeung in \cite{Cai2004} and Devetak in \cite{Devetak2005}\cite{Devetak2005-1}.

On the one hand, QKD with satellites is extremely challenging due to the small signal strength of the order of one photon per pulse, the high channel loss, delays due to the iterative protocol, and the sensitivity to background noise, which does not allow key exchanges during daytime. On the other hand, eavesdropping without being noticed seems to be very complicated as well in a free-space scenario. So it makes a lot of sense to consider a few reasonable assumptions which limit the power of Eve. Indeed, due to the laws of celestial mechanics a satellite cannot be parked in the line of sight between Alice's satellite and Bob's ground-station (or satellite), which prevents intercept and resend attacks. Eve could  position her satellite very close or eventually attach it to Alice's satellite. This would allow Eve to exploit side channels (e.g. monitoring electronic signals), but this scenario is excluded in QKD which relies by definition on secure offices. Similarly, Eve could place a telescope close to Bob's but again we may introduce an exclusion region under control of Bob, within which Eve cannot install a big telescope. 

There is some initial work investigating the consequences of restricted power for Eve on the performance of QKD links \cite{Shapiro2019}\cite{Shapiro2019-1}\cite{Ling2019}\cite{Fujiwara2018}. However, given their additional, reasonable assumptions, we find exactly the scenario where keyless private communication is possible. This type of secret transmission and its practical comparison with QKD have been studied in \cite{Endo2015}\cite{Endo2016}, however, without  assuming a quantum channel. 

In this paper, we show that physical layer encryption is a reasonable choice for satellite communication over the quantum space channel. In Section II, we introduce the one-way wiretap protocol and identify its asymptotic information-theoretical secure and reliable rate: the private capacity. In Section III, we propose a simple yet accurate channel model using binary modulated coherent states (on-off keying, OOK) and derive and analyze the private capacity for Bob using realistic, state-of-the-art photon counting detectors while the signal reception by Eve being only limited by the laws of physics. Finally, we present numerical results for a realistic scenario taking the performance of the Micius system as a reference and compare it to the fundamental limit for the secret key rate of QKD \cite{PLOB2017}.

\section{Description of the protocol}

Our protocol is based on the one-way wiretap protocol \cite{Wyner1975}\cite{Csiszar1978} where secret bits are channel encoded and sent over $n$ uses of the optical channel. We assume OOK for lighter analytical treatment, but other binary modulations can be used e.g. Binary Phase Shift Keying (BPSK). 

The protocol contains the following steps.
\begin{enumerate}

\item \textbf{Encoding.} Alice, the legitimate information transmitter, sends a stream of secret information bits (any data, video, pictures, etc.) to a stochastic wiretap encoder. Information-theoretically, this encoding is described as Alice selects a codeword $X^n$ to send her secret message $M$. The signal is meant for Bob, but part of it is leaked to the environment represented by the omnipresent Eve. The secrecy depends on the structure of this encoder, 
which is characterized by the rate $R=k/n$ (where $k$ is the number of secret bits), the error probability, $\epsilon_n$, and a security measure, $\delta_n$.

\item \textbf{State Preparation.} For each use of the channel, the legitimate information transmitter (Alice), prepares a coherent state modulated by the random variable $X \in \mathcal{X}=\{0,1\}$, where $X=0$ with probability $q$ and $X=1$ with probability $1-q$. The OOK states transmitted by Alice are the vacuum state, $\ket{\alpha_0} = \ket{0}$, and
\begin{align}
\ket{\alpha_1} &= e^{-\frac{1}{2}|\alpha_1|^2} \sum_{n=0}^{\infty} \frac{\alpha_1^n}{(n!)^{1/2}}\ket{n}.
\end{align}
The probability $q$ is not decided \emph{a priori} as part of the protocol, it needs to be optimized depending on the assumptions at the detection and the physical propagation channel (Section III).

\item \textbf{Measurement.} After $n$ transmissions over the quantized propagating field, Bob receives $B^n$ and Eve $E^n$. Bob estimates his received coherent state after measuring the arriving light and obtains $Y^n$. He disposes of state-of-the-art detectors. Eve may apply the best quantum detection strategy to obtain $Z^n$.

\item \textbf{Decoding.} Bob and Eve send their estimated received states to the decoder. The concrete construction of encoder and decoder are assumed to be publicly known. The  values of $\epsilon_n$ and $\delta_n$ depend on the choice of this code.
 
\end{enumerate}

According to the wiretap theory, even if the eavesdropper is computationally unbounded, the wiretap code ensures that if $R$ is an achievable rate, both $\epsilon_n$ and $\delta_n$ (after decoding) tend to zero for large $n$,
\begin{align}
\lim_{n \rightarrow \infty} \epsilon_n = \lim_{n \rightarrow \infty} \delta_n &= 0,
\end{align}
which means that the error probability and the information leakage towards Eve can be made arbitrarily low. Upper bounds are also known for the speed of convergence rate of leaked information \cite{Hayashi2015}. Hence, the protocol jointly provides information-theoretical reliable and secure communication. The supremum of all such rates is called the private capacity and for the classical-quantum wiretap channel subject to the strong security criterion, has been determined in \cite{Cai2004}\cite{Devetak2005}. The meaning of strong security is that, given a uniform distribution of the message to be transmitted through the channel, the eavesdropper shall obtain no information about it \cite{Maurer1994}. This criterion is the most common security criterion in classical and quantum information theory. The metric of strong security is the amount of mutual information leaked to Eve.

When Bob's channel is degradable, the private capacity of the quantum wiretap protocol (with quantum channel and information) is \cite{Devetak2005-1}\cite{Smith2008}\cite{Wilde2018}
\begin{align}
C_P(\mathcal{N}) =  \sup_{\rho_{in}} I_c(\rho_{in},\mathcal{N}), 
\label{PriCapFormula}
\end{align}
where $\rho_{in}$ is the input density matrix and $ I_c(\rho_{in},\mathcal{N})$ is the coherent information defined as 
\begin{align}
I_c(\rho_{in},\mathcal{N})= S(\mathcal{N}_{B}(\rho_{in})-S(\mathcal{N}_{E}(\rho_{in})), 
\end{align} \label{eq:CoherentInfo}
with $S$ the von Neumann entropy. In the next sections we characterize the degradable channel of our practical (energy-constrained) protocol over space links, which we use to derive the private capacity.
\begin{figure}[tbh]
\centering
\includegraphics[scale=0.2]{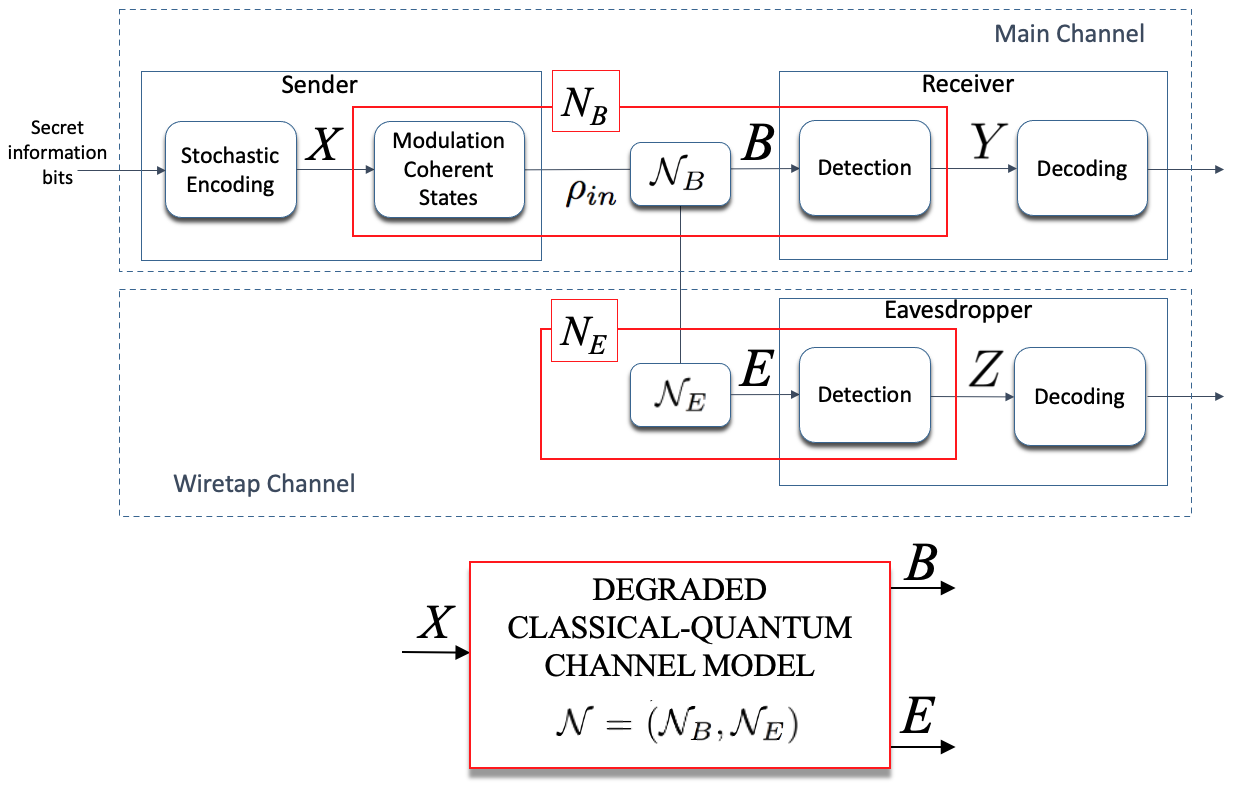}
\protect\caption{The main channel and the wiretap channel (top) and the information-theoretical representation of the degraded classical-quantum channel (bottom).} 
\label{fig:sysmodel}
\end{figure}

The (physical) description of our protocol is illustrated in Fig. \ref{fig:sysmodel} (top). The figure shows encoded bits modulating coherent states, the quantum channel $\mathcal{N}=(\mathcal{N}_B,\mathcal{N}_E)$, the resulting classical channel after detection $N=(N_B,N_E)$ as well as the information theoretical representation (bottom).

\section{Calculation of the private capacity $C_{P}$}

First, we build a model for the classical-quantum channel of our protocol. The legitimate transmitter, Alice, is on a satellite while the legitimate receiver, Bob, is on Earth. We assume a single-mode free-space quantum bosonic channel (for the wiretap channel in the semi-classical regime see \cite{Hayashi2020}).  The overall efficiency of Bob's channel is $\eta$. The coefficient $\gamma \in (0,1)$ characterizes the channel degradation. Hence, the efficiency of Eve's channel is $\gamma\eta$.

According to step 2 of our protocol, the received states are simply the vacuum, or $\ket{\sqrt{\eta}\alpha_1}$  and $\ket{\sqrt{\eta\gamma}\alpha_1}$ for Bob and Eve respectively. The wiretap channel transition probabilities depend on the coherent states received by Bob and Eve and by their detection strategies. For practical purposes, we assume Bob uses standard single photon detectors, i.e. a threshold detector. 
For Bob, we also take into account limited detection efficiency (included in $\eta$, see Appendix \ref{sec:App4}) and noise (dark counts probability $p_{dark}$ and stray light with a Poisson photon number distribution and average $\eta_0\Delta$). The conditional probabilities that Bob detects $y$ given that Alice sent $x$ are illustrated in Fig. 3 with $\epsilon_0=(1-p_{dark})e^{-\eta_o\Delta}$ and $\epsilon_1=(1-p_{dark})e^{-(\eta\mu +\eta_o\Delta)}$, where we have denoted $\mu=|\alpha_1|^2$. 

Eve instead performs an optimal quantum detection. For the single observation, this leads to the optimal error probability $\epsilon^*$, which can be calculated as \cite[Lemma 3.1]{Hayashi2006}\cite{Audenaert2007}\cite{Audenaert2008}, 
\begin{align}
\epsilon^* &=\min_{0 \leq  \Pi \leq \mathbf{1}} [ q\Tr (\mathbf{1} - \Pi)\rho_1 + (1-q)\Tr \Pi \rho_0 )] \\
&= \frac{1}{2}-\frac{1}{2}\norm{q\rho_1 - (1-q) \rho_0}_1,
\end{align} \label{eq:minPe}
where $\norm{}_1$ stands for trace norm. In the binary source scenario, this bound is achieved by the Holevo-Helstrom projector 
\cite{Helstrom1976}\cite{Holevo1978}, for which several practical implemetations have been proposed such as the Kennedy receiver \cite{Kennedy1973}, the Dolinar receiver \cite{Dolinar1973} or the Sasaki-Hirota receiver \cite{Takeoka2008}\cite{Tsujino2011}. The optimal error probability of Eve resulting from (5), becomes
\begin{align}
\epsilon^*(\gamma)=(1-\sqrt{1-4q(1-q)e^{-\eta\gamma \mu}})/2. 
\end{align}

\begin{figure}
\centering
\includegraphics[width=3.3in]{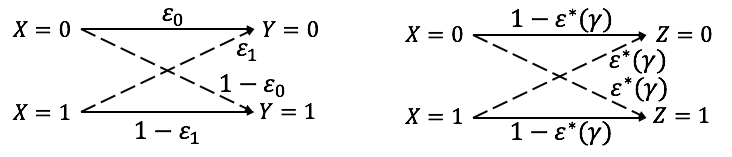}
\caption{Probabilistic detection models assumed for Bob and Eve to obtain $Y$ and $Z$, respectively.}
\label{fig:Degradation}
\end{figure}

The private capacity for our wiretap channel model coincides with the classical secrecy capacity and is defined as
\begingroup
\small
\begin{align}
C_P(\gamma)&= \max_q \{ I(X;Y) - I(X;Z | \gamma ) \},
\label{eq:PrivateRate}
\end{align}
\endgroup
where $I(X,Y)$ is the Shannon mutual information of Alice using a state-of-the-art photon counting detector and $I(X;Z | \gamma )$ is the maximum Shannon mutual information that Eve can physically detect. For our setting as presented in Section IV, a uniform input probability (i.e. $q=0.5$) is very close to optimum (see Appendix \ref{sec:App0}), and we have
\begingroup
\small
\begin{align}
C_P(\gamma)&= \left[h(\epsilon^*(\gamma)) + h\left(\frac{\epsilon_0+\epsilon_1}{2}\right) - \frac{h(\epsilon_1)+h(\epsilon_0)}{2} - 1\right]_+,
\end{align}
\endgroup

where $[]_+$ means the positive part and we have used the notation $h()$ as the binary Shannon entropy. If we now restrict Eve only by the laws of quantum electrodynamics, we obtain the Devetak-Winters rate \cite{DW2005} for our protocol

\begin{align}
R_{DW} (\gamma)
& = I(X,Y)  - \mathcal{\chi}(X;E|\gamma),
\end{align}
where $I(X;Y)$ is the Shannon mutual information of Alice's choice for input probability and measured by a
state-of-the-art photon counting detector. The quantity $\mathcal{\chi}(X;E|\gamma)$ is the Holevo bound for the eavesdropper. Denoting $\epsilon(\gamma)=\bra{0}\ket{\sqrt{\eta\gamma}\alpha_1}$, for uniform priors (i.e. $q=0.5$) it is easy to show

\begingroup
\small
\begin{align}
R_{DW} (\gamma)
&= \left[h\left(\frac{\epsilon_0+\epsilon_1}{2}\right) - \frac{h(\epsilon_1)+h(\epsilon_0)}{2} - h\left(\frac{1+\epsilon(\gamma)}{2}\right) \right]_+,
\label{eq:DWPrivateRate}
\end{align}
\endgroup

\begin{figure}[tbh]
\centering
\includegraphics[scale=0.2]{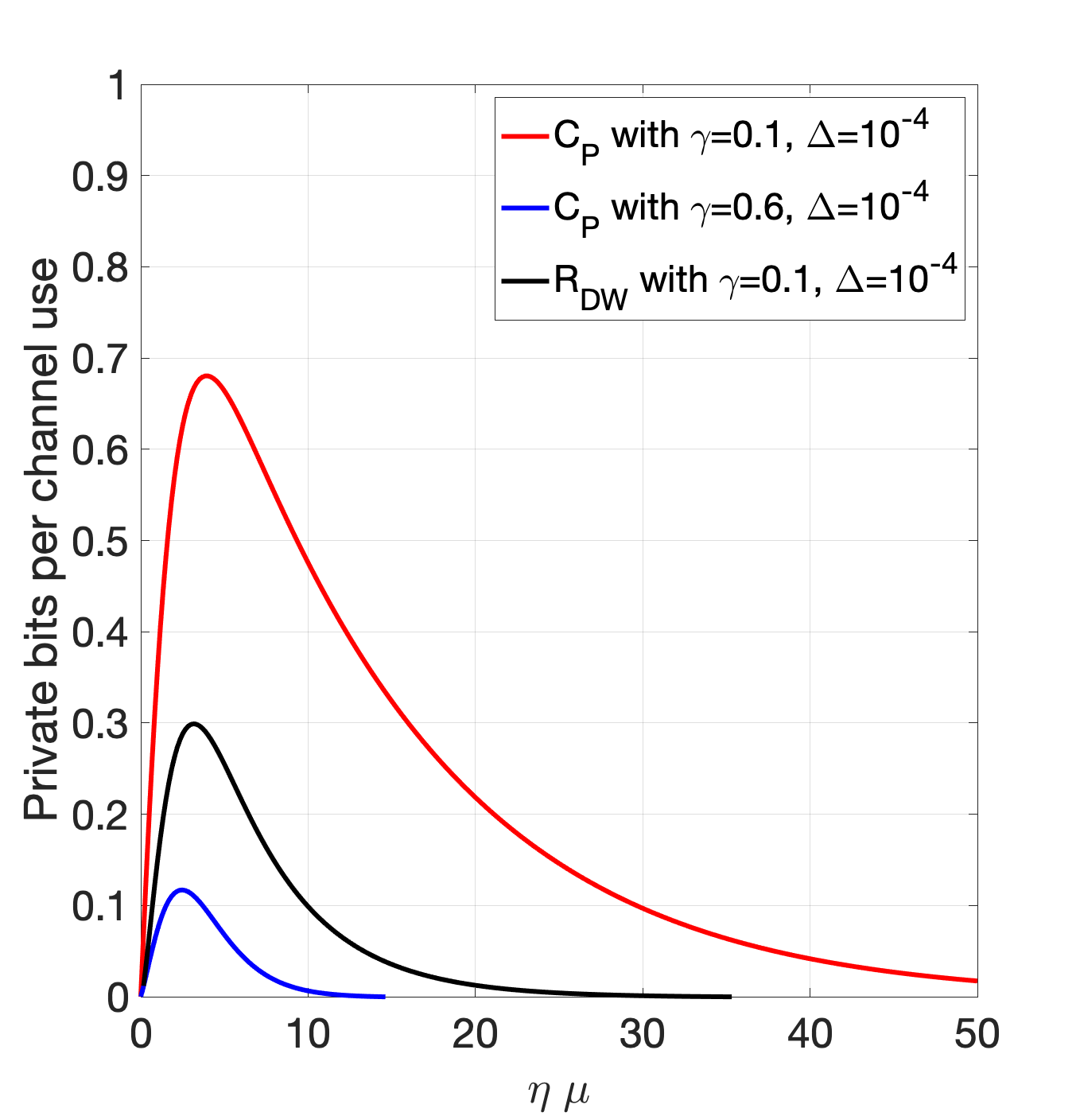}
\protect\caption{Private capacity of OOK for $\gamma=0.1$ and $\gamma=0.6$ ($p_{dark} \approx 0$, $\eta_o=1$  $\Delta=10^{-4}.$) For comparison, also the numerical values of $R_{DW}$ for $\gamma=0.1$. }
\label{fig:Copt}
\end{figure}

The numerical calculation is shown in Fig. \ref{fig:Copt}. We observe that the private rate in \eqref{eq:PrivateRate} is as high as 0.68 for $\gamma = 0.1$ (with little sensitivity to noise, see Appendix  \ref{sec:App1}). And even for adverse channel conditions ($\gamma$ close to 1) there is a positive secrecy rate for reasonable noise (e.g.up to $\gamma = 0.6$ for $\Delta = 10^{-4}$), however we will show below that in practice is best to aim for $\gamma < 0.1$. For comparison, we have also plotted \eqref{eq:DWPrivateRate} for $\gamma=0.1$. We observe a substantial decrease, which is also affected by noise, however, we still get significant positive rates. In the following, we assume the realistic setting of Eve using Helstrom detection.

\section{The Micius satellite: a concrete example of an orbit-earth optical link}

\begin{figure}[tbh]
\centering
\includegraphics[scale=0.4]{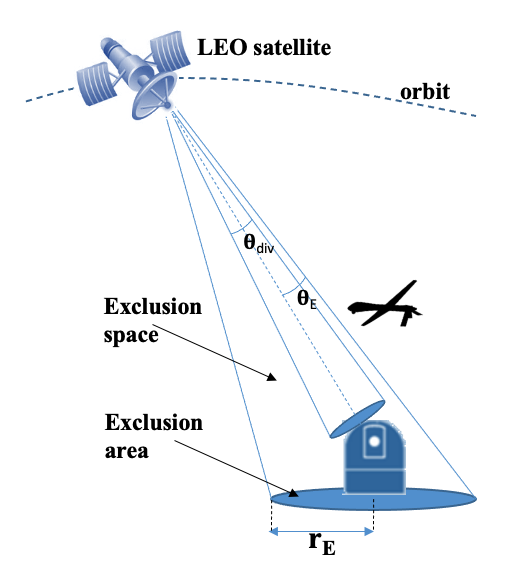}
\protect\caption{Geometry for the secure and reliable communication of our wiretap protocol showing the exclusion radius $r_E$.} 
\label{fig:Geometry}
\end{figure}

As a realistic physical scenario, we use as a reference the recent experiment of QKD with the Chinese LEO satellite Micius \cite{Micius2017}\cite{Micius2020}. The geometry is shown in Fig. \ref{fig:Geometry}. The satellite has an orbit of about 500 km above the earth surface and exchanges keys over distances up to 1200 km if the satellite is close to the horizon. The transmitter is equipped with 300 mm Cassegrain telescope featuring a far field divergence $\theta_{div}$ of 10 $\mu$rad (full angle at $ 1/e^2$). The receiver at groundstation has a telescope with a diameter $D_R$ of 1m. 

We can approximate the fraction of the light collected by Bob, the free space loss $\eta^B_{f}$, as the ratio of the telescope area and the footprint area
\begin{equation}
\eta^B_{f}=\frac{A_{R}}{A_{F}}=\dfrac{\pi (D_R/2)^2}{\pi (D_F/2)^2}=\frac{D_R^2}{\theta_{div}^2d^2_B},
\label{eq:BobBudget}
\end{equation}
with $$D_F \approx \theta_{div}  d$$
where  $d_B$ is the distance between transmitter and receiver. 

The number of photons detected by Bob is given by 
$\eta \mu =\eta^B_{f}\eta_{b} \mu,$ where  $\eta_{b}$ is the total additional loss of Bob depending on the experimental situation. In the Micius experiment those are atmospheric turbulence 3-8 dB ($\eta_{atm}$), pointing errors ($\eta_p$) $<$ 3 dB, overall optical loss ($\eta_o$) from telescope input lens to detector 7.4 dB detector, detector efficiency $\eta_{det}$  50 \% (3 dB) (see Appendix \ref{sec:App4}). In the following we calculate with a loss $\eta_b$ of 20 dB (1\%). 

For Eve we calculate $\eta^E_{f}$ as in (\ref{eq:BobBudget}) but we add a factor taking into account the light intensity outside the exclusion angle supposing a Gaussian angular distribution of the beam as 

\begin{equation}
\eta^E_f=\frac{(D^E_R)^2}{\theta_{div}^2d_E^2} e^{-2(\frac{2\theta_E}{\theta_{div}})^2}.
\end{equation}

Then, the number of photons detected by Eve becomes simply $\eta^E_f \mu=\gamma \eta \mu$, as we assume no additional loss for Eve. Hence, for fixed antenna sizes we can easily calculate $\gamma$ as 

\begingroup
\small
\begin{align}
\gamma(d_B, d_E,\eta_b, \theta_E, \theta_{div})
&=\dfrac{1}{\eta_b}\Big( \frac{d_B}{d_E} \Big)^2\Big( \frac{D^E_R}{D^B_R} \Big)^2 e^{-2(\frac{2\theta_E}{\theta_{div}})^2}.
\label{eq:physgamma1}
\end{align}
\endgroup

For $d_B=d_E= 1200~km$ and $\theta_E=r_E/d$, for the Micius system parameters and assuming a very large eavesdropper's receiving antenna $D^E_R$ of 2 m and a small exclusion radius  $r_E=12.5$m we obtain: 
$$\gamma = 0.07<0.1$$
For the reminder, we fix the exclusion radius such that $\gamma <0.1$. Indeed, this value seems to be a good choice as it leads to high secret capacities $>$ 0.6 (see Fig. \ref{fig:Copt}), little sensitivity to noise and signal fluctuations for reasonable exclusion radii as discussed in Appendix \ref{sec:App1}. This sensitivity is driven by the distinguishability of the coherent states at Eve's Holevo-Helstrom detector as shown in Appendix \ref{sec:App5}, the lower the $\gamma$ the less sensitivity of the distinguishability to signal dynamics.

\section{Private rates for different practical settings and comparison with QKD}


\renewcommand{\arraystretch}{1.8}
\begin{table*}[ht]
\centering
\resizebox{\textwidth}{!}{
\begin{tabular}{|c|c|c|c|c|c|c|c|}
\hline 
\multirow{2}{*}{Configuration } & \multirow{2}{*}{Distance [km]} & \multirow{2}{*}{Channel loss ($\eta^B_f$)} & \multicolumn{2}{c|}{QKD (night) $\Delta = 10^{-7}$} & \multicolumn{3}{c|}{Wiretap channel (day) $\Delta = 10^{-4}$} \\ \cline{4-8} 
 &  &  & Micius & PLOB ($\eta_b = 1$) & exlusion radius $r_E$ [m] & gamma &  private rate \\ 
\hline 
LEO & 500 - 1200 & 22dB & $<$ 10kHz & 10MHz & 12.5 & 0.1 & 700MHz \\ 
\hline 
MEO & 10000 & 40dB & - & 100kHz & 100 & 0.1 & 700MHz \\ 
\hline 
GEO & 36000 & 52dB  & - & 6kHz & 340 & 0.1 & 700MHz \\ 
\hline 
\end{tabular}}
\caption{Comparison of the achievable secret key rates for QKD and the private rates for the wiretap channel presented in this work. The values are extrapolated from experimental data from the Micius satellite [4] for  $\eta^B_f=22dB$ (@1200 km) and $\eta_b = 20dB$ (except for the theoretical PLOB bound with $\eta_b=1$). The exclusion radii $r_E$ for $\gamma=0.1$ are calculated based on the Micius beam parameters. Note that for Micius and PLOB we consider nighttime operation, in contrast to the case of the wiretap channel where daytime operation is possible.} 
\end{table*}

We can now calculate the private capacity for different geometrical configuration, supposing that Alice and Bob have a satellite and a ground station equivalent to the Micius experiment. We consider OOK with a clock rate of 1GHz. With a time window of 1 ns, state of the art single photon detectors feature a $p_{dark} < 10^{-7}$, so detector noise has no significant effect on the secret capacity.  Table 1 in \cite{Er-lon2005} indicates the expected number of noise photons for different collection angles, filter bandwidths and temporal windows. Based on that an average number $\Delta$ of $10^{-4}$ and $10^{-7}$ is an achievable value for clear daytime sky, and a full moon clear night, respectively. During a cloudy day, one could expect a $\Delta$ of $10^{-2}$, and still positive private rates (this is under the assumption that transmission of the channel is affected in the same way for Bob and Eve). See Figure \ref{fig:PvsG} for the sensitivity to noise.  

Given the clock rate and the maximum private capacity from Fig. \ref{fig:Copt}, we obtain a private rate of $680Mb/s$ for a $\gamma=0.1$. This value is the same for any channel efficiency $\eta$ as long as the received power $\eta\mu $ is optimized. We can compare these rates with the key rates obtained by QKD in the Micius experiment and a potential perfect quantum communication scheme. Indeed, Pirandola et. al. showed recently \cite{PLOB2017} that there is an upper bound of the achievable secret key rate of a QKD protocol for a given channel transmission $\eta$
\begin{align}
R_{QKD,\infty}(\eta)= -\log_2(1 - \eta).
\end{align}

Table I presents all these rates for LEO, MEO and GEO satellites. As expected, the private capacity of a wiretap channel outperforms QKD dramatically in terms of rate and most importantly in terms of resistance against noise. 

Certainly, if the eavesdropper is restricted also the performance of QKD can be improved substantially. As a comparison, we estimate the theoretically achievable secret key rate for QKD following Appendix B in \cite{Shapiro2019}, choosing the parameters such that they are equivalent to our assumptions, notably $\Delta=10^{-4}$ and $\gamma=0.1$. We obtain a similar rate of 360 Mbits/s. However, keyless private communication is arguably easier to implement in practice, in particular it doesn't need two way communication for reconciliation.

There is also a scheme called Quantum Secure Direct Communication \cite{Qi2019}, which combines quantum communication and the possibility to detect the presence of an eavesdropper with encoding like for the wiretap channel. However, since this scheme is impractical, we don't consider it in this comparison. 

Finally, some practical remarks. The necessary lasers power in order to reach the optimal signal strength of about 4 photons in average is about 15 mW and 15$\mu$W for the GEO and the LEO setting, respectively, and therefore it is no limitation. However, due to the high signal strength, a GHz repetition rate implies high detection rates at Bob's, which are still a challenge for single photon detectors \cite{matthieu20}.

\section{Discussion and Conclusions} 

We have shown the feasibility, practicality and performance of unconditionally secure links for space and show the required exclusion radius. A protection area is needed for any kind of secure communication, including QKD. In this paper, we discussed a downlink, however, similarly, we could also consider an uplink and estimate the channel degradation $\gamma$ for reasonable assumptions on Eve's satellites. 
 
Our protocol is certainly susceptible of jamming attacks, but so are QKD protocols. However, while our protocol cannot protect the communication from jamming and active attacks, it can be used in coordination with security mechanisms in communication layers above the physical layer to provide the satellite system availability, integrity and confidentiality. 

Given these boundary conditions, we have demonstrated that physical layer encryption can provide information theoretically secure communication also in the presence of Eve only limited by the laws of quantum physics. As for the wiretap codes, explicit constructions are available that can provide the strong security. Well known constructions are based either on coset codes with random codeword choice within a coset \cite{Thangaraj2007}\cite{Klinc2011} or concatenation of encryption functions (e.g. hashing) with conventional capacity-achieving codes \cite{Tyagi2015}\cite{Hayashi2020} (e.g. low-density parity code (LDPC), polar code or Reed Muller codes).

The achievable private rates are considerably higher than the QKD rates for an unrestricted Eve. Moreover, direct private communication is also possible close to illuminated cities and even during daytime in contrast to QKD. 
Moreover, given the low rates, the secret keys generated by QKD will in practice not be used in combination with the one-time-pad but with symmetric encryption systems like Advanced Encryption Standard (AES). This means that the legitimate users have to choose between trusting physical security including exclusion areas around Alice and Bob (which is arguably needed for QKD as well) or the computational security of encryption algorithms. 
 
Overall, physical layer encryption seems to be a reasonable choice for satellite communication.

\begin{acknowledgments}
We would like to acknowledge Antonio Ac\'in, Andreas Winter, Masahito Hayashi, Christoph Wildfeuer and Mikael Afzelius for useful discussions. D.R. acknowledges the support of European Union’s Horizon 2020 program under the Marie Skłodowska-Curie project QCALL (No. GA 675662).
\end{acknowledgments}

\appendix
\counterwithin{figure}{section}

\section{\label{sec:App0} Optimized input probabilty}
\begin{figure}[tbh]
\centering
\includegraphics[scale=0.45]{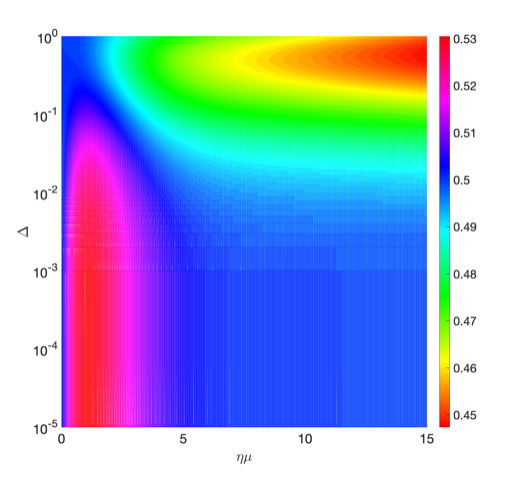}
\protect\caption{Optimal value of input probability, $q$, as a function of the average number of photons received by Bob $\eta\mu$ and the intensity of the stray-light $\Delta$ ($\gamma=0.1$).} 
\label{fig:optq}
\end{figure}

The optimization of the private capacity given in \eqref{eq:PrivateRate} also provides the optimal input probability that Alice should choose to transmit the OOK symbols. Fig. \ref{fig:optq} shows the optimized input probability as a function of the received average number of photons and intensity of stray light. It is interesting to observe that our wiretap channel can be considered nearly symmetric. However, optimal private rates require small deviations from the symmetric channel.

\section{\label{sec:App1} The appropriate value of $\gamma$ and the corresponding exclusion radius}

In this section we give a justification for the value for $\gamma=0.1$ as a good trade-off between, on the one hand, a high secrecy capacity, and on the other hand, the required exclusion area and resilience to experimental fluctuations.

\begin{figure}[tbh]
\centering
\includegraphics[scale=0.4]{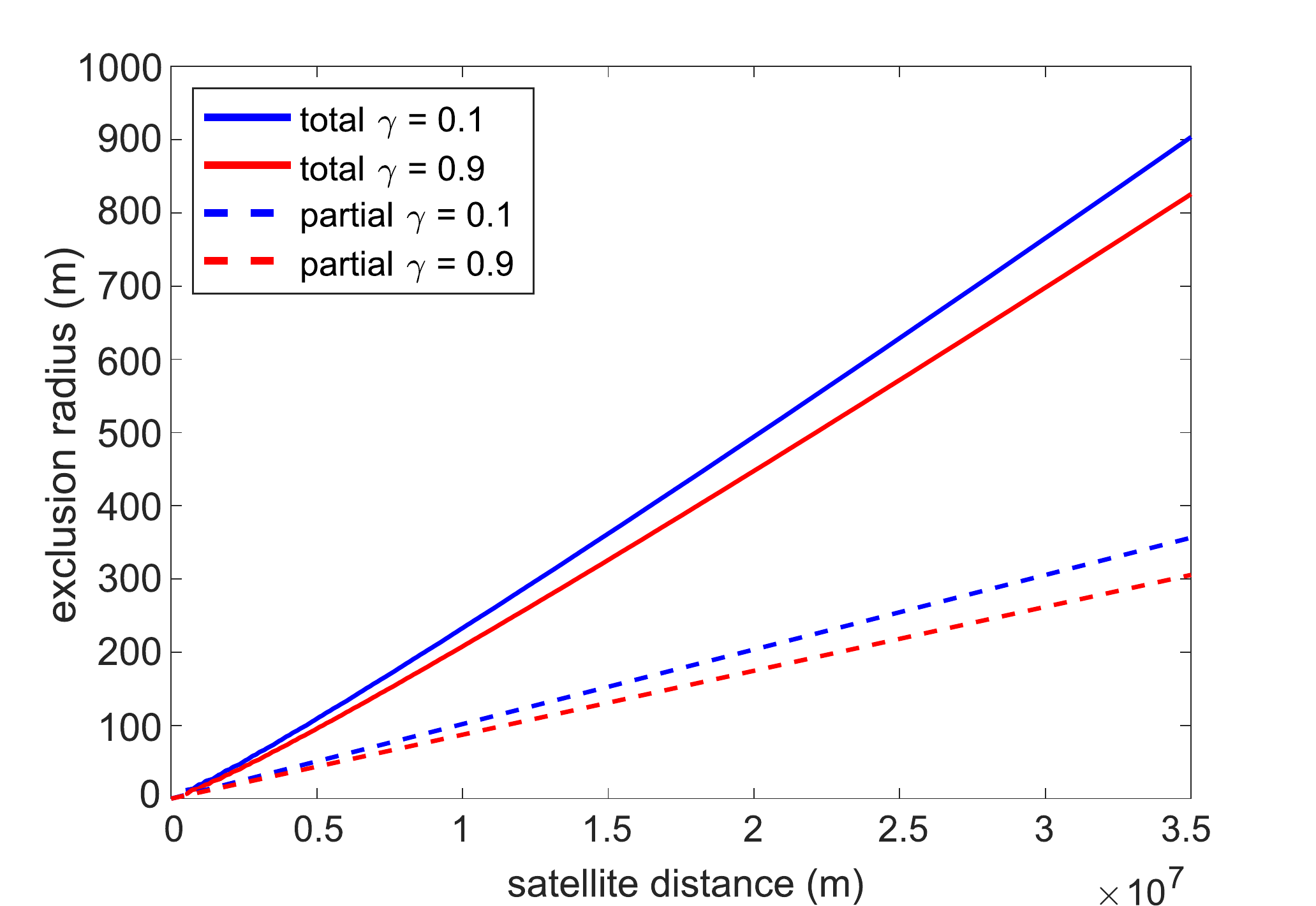}
\protect\caption{Exclusion radius vs. the satellite distance for different $\gamma$ and eavesdropper model. "Partial" stands for a telescope of $2m$ of diameter, "total" for an eavesdropper collecting all light outside the exclusion region.} 
\label{fig:radius}
\end{figure}

We consider the situation in which the adversary Eve is going to collect the light with a telescope from a location close to Bob, hence at the same distance $d$ from the satellite, Alice. In this case, if $\gamma$ is fixed, the value of the exclusion ratio is given by the formula:

\begin{equation}
r_{ex} = \frac{1}{2}\theta_{div}d\sqrt{\left(\frac{1}{2}log\left(\frac{1}{\gamma}\frac{1}{\eta_b}\left(\frac{D^E_R}{D^B_R}\right)^2\right)\right)}
\end{equation}

where $\theta_{div}$ is the far field divergence angle for the Gaussian beam, $\eta_b$ is the efficiency of the detection system of Bob and $D^E_R$ and $D^B_R$ are respectively the radius of the antennas of Eve and Bob respectively.

As it can be seen from Fig.~\ref{fig:radius} the requested exclusion radius depends linearly on the satellite to ground distance and it amounts to about $350~m$ for  a geostationary orbit (around $35000~km$) for a $\gamma = 0.1$. It can be also seen that increasing the value of $\gamma$, e.g. $\gamma = 0.9$, doesn't allow us to reduce the exclusion radius significantly.  

A more restrictive approach consist in giving Eve all the optical power outside the exclusion radius. This correspond to assuming that Eve can collect all the light outside the cone from the satellite to the secure area. In this case the exclusion ratio can be extrapolated from the following implicit function once $\gamma$ is fixed:

\begin{equation}
e^{-2\left(\frac{r_{ex}}{\theta_{div}d}\right)^2} = \gamma\left( 1-e^{-2\left(\frac{D^B_R}{\theta_{div}d}\right)^2}\right)
\end{equation}

\begin{figure}[tbh]
\centering
\includegraphics[scale=0.4]{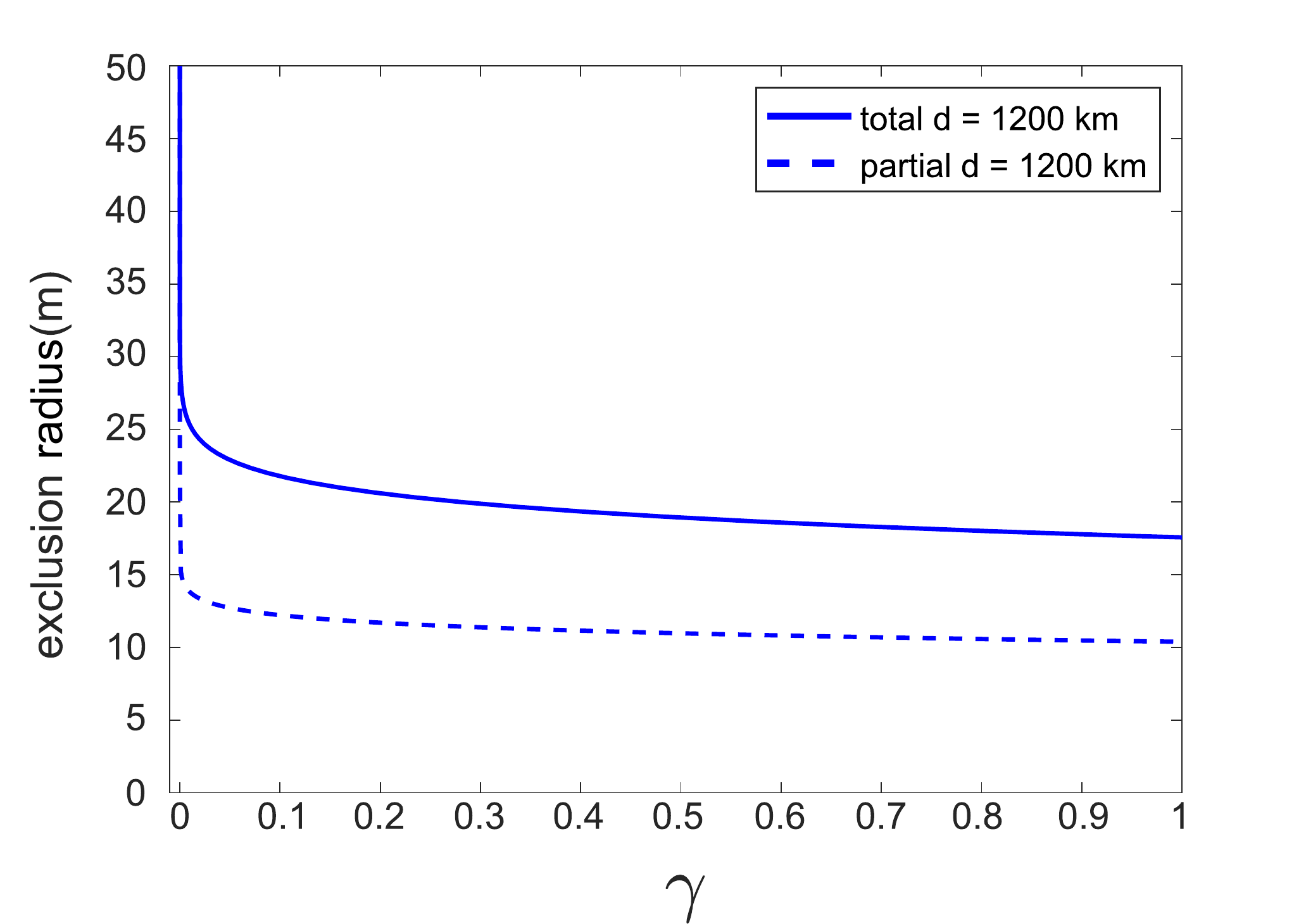}
\protect\caption{The exclusion radius needed to achieve a given value of $\gamma$. }
\label{fig:radius_g}
\end{figure}

Also in this case the relation between the exclusion radius and the satellite to ground distance is almost linear. However, the value is two to three times higher than in the previous scenario. Still, the exclusion radius for the geostationary orbit remains below $1~km$.

Fig.~\ref{fig:radius_g} shows the dependence of the exclusion radius on the parameter $\gamma$ when the distance satellite to ground is fixed to the value of 1200 km (distance similar to the Micius satellite). It illustrates that indeed (for a finite detection area as well as for the unlimited case) the required exclusion radius is almost constant with respect of $\gamma$ for values higher than $0.1$. This means that going for value for $\gamma$ higher than $0.1$ brings only a little advantage in terms of exclusion radius, but at the cost of a significant reduction of the secrecy rate, detrimental for the performance of the protocol. Moreover, in this range, a small uncertainty in the radius, or a bad pointing stability, may lead to a large variation in $\gamma$ and threaten the security of the protocol. Therefore, going for values of $\gamma<0.1$ seems to be pertinent, in the case of optical communication with well defined Gaussian beams. 

\begin{figure}[tbh]
\centering
\includegraphics[scale=0.4]{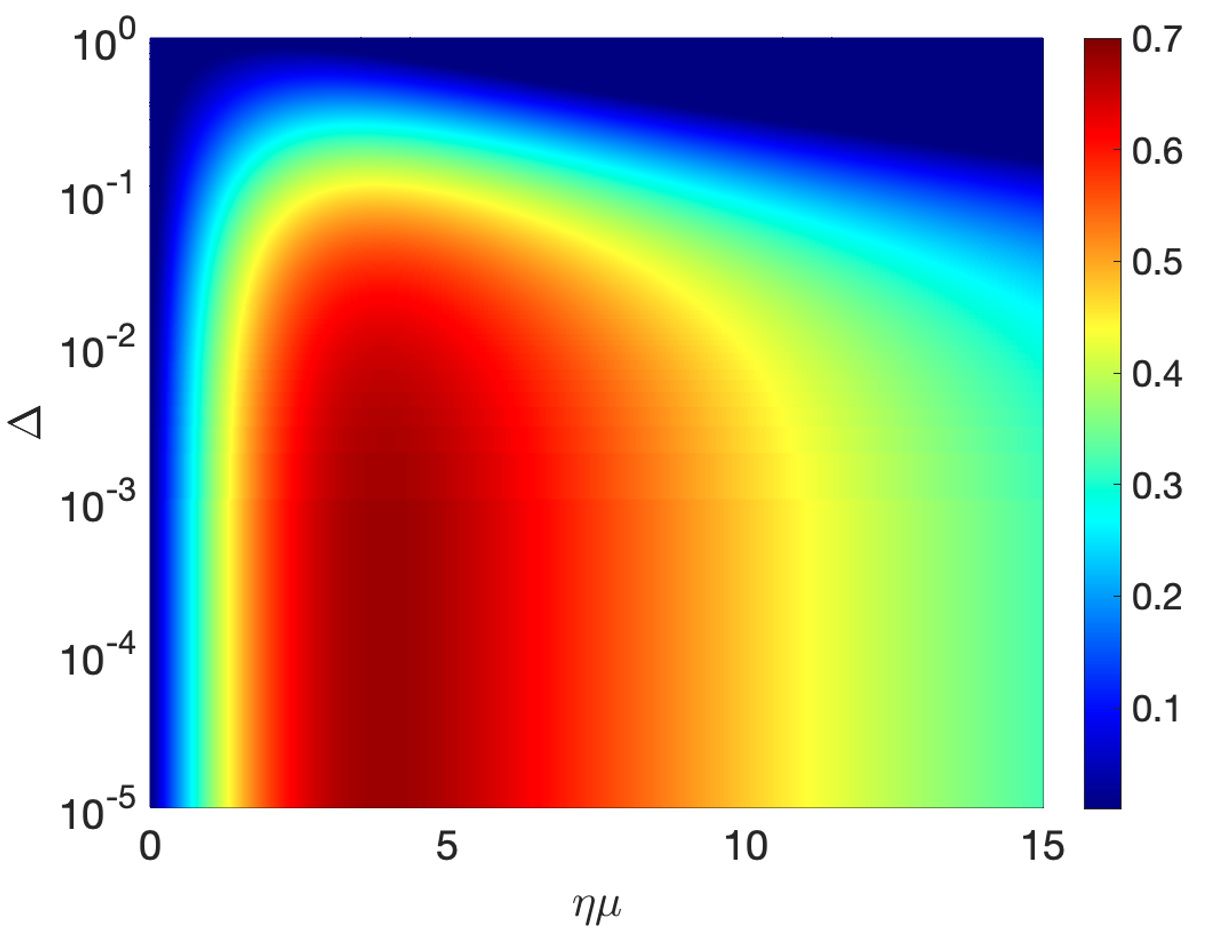}
\protect\caption{Private capacity $C_p$ for $\gamma=0.1$ as a function of the average number of photons received by Bob and the intensity of the stray-light.}
\label{fig:PvsG}
\end{figure}

Finally in Fig.~\ref{fig:PvsG} we show the behaviour of the private capacity $C_p$ for $\gamma$ equal to $0.1$, with respect to stray light  $\Delta$ and the received light by Bob $\eta\mu$ (both in units of average number of photons per time and frequency bin). We note that if we fix a secret capacity of say 0.5, the protocol can resist to high fluctuations in the received light intensity and a high amount of stray light (remember, we expect a $\Delta$ of $0.0001$ for a clear sky during daytime). This security margin would be considerably reduced for a $\gamma$ of 0.6 for instance.

\section{\label{sec:App3}: Basic celestial mechanics}

\begin{figure}[tbh]
\centering
\includegraphics[scale=0.45]{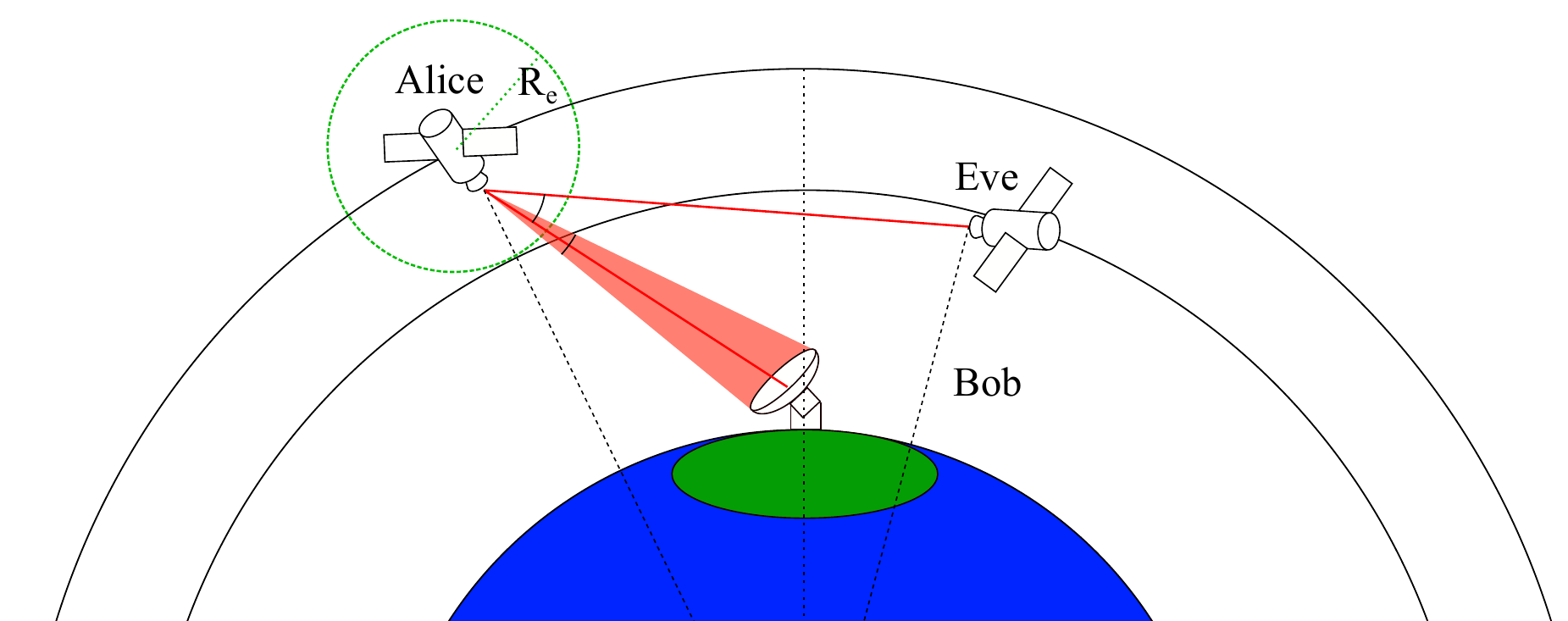}
\protect\caption{Sketch of the orbits of Alice and Eve with respect to Bob.} 
\label{fig:SatAtt}
\end{figure}

Since, as discussed above, Eve can get only limited information by positioning a receiver station on the ground next to Bob's telescope, the natural question would be what happens if Eve tries to attack Alice satellite instead? In this section we give some arguments that, similarly to the previous case, it is sufficient to assume a reasonable exclusion radius around Alice satellite in order to assure the security of our protocol. A complete treatment of the question is out of the scope of this paper, we just present some simple, general considerations. (see for e.g. \cite{Tapley2004} for more information )

\begin{figure*} 
\centering
\begin{subfigure}{.45\textwidth}
  \centering
  \includegraphics[width=1\linewidth]{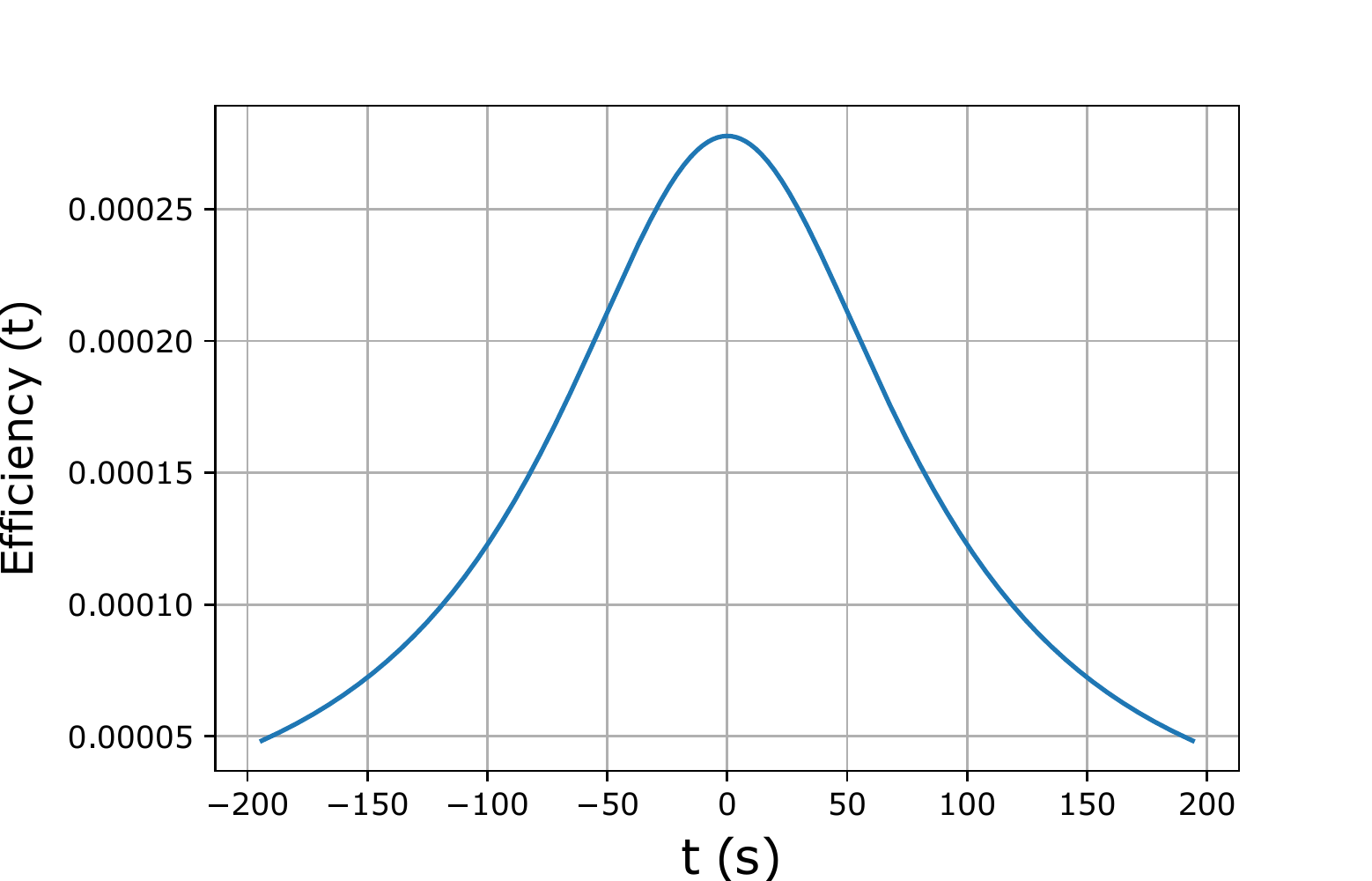}
  \caption{}
  \label{fig:sub1}
\end{subfigure}%
\begin{subfigure}{.45\textwidth}
  \centering
  \includegraphics[width=1\linewidth]{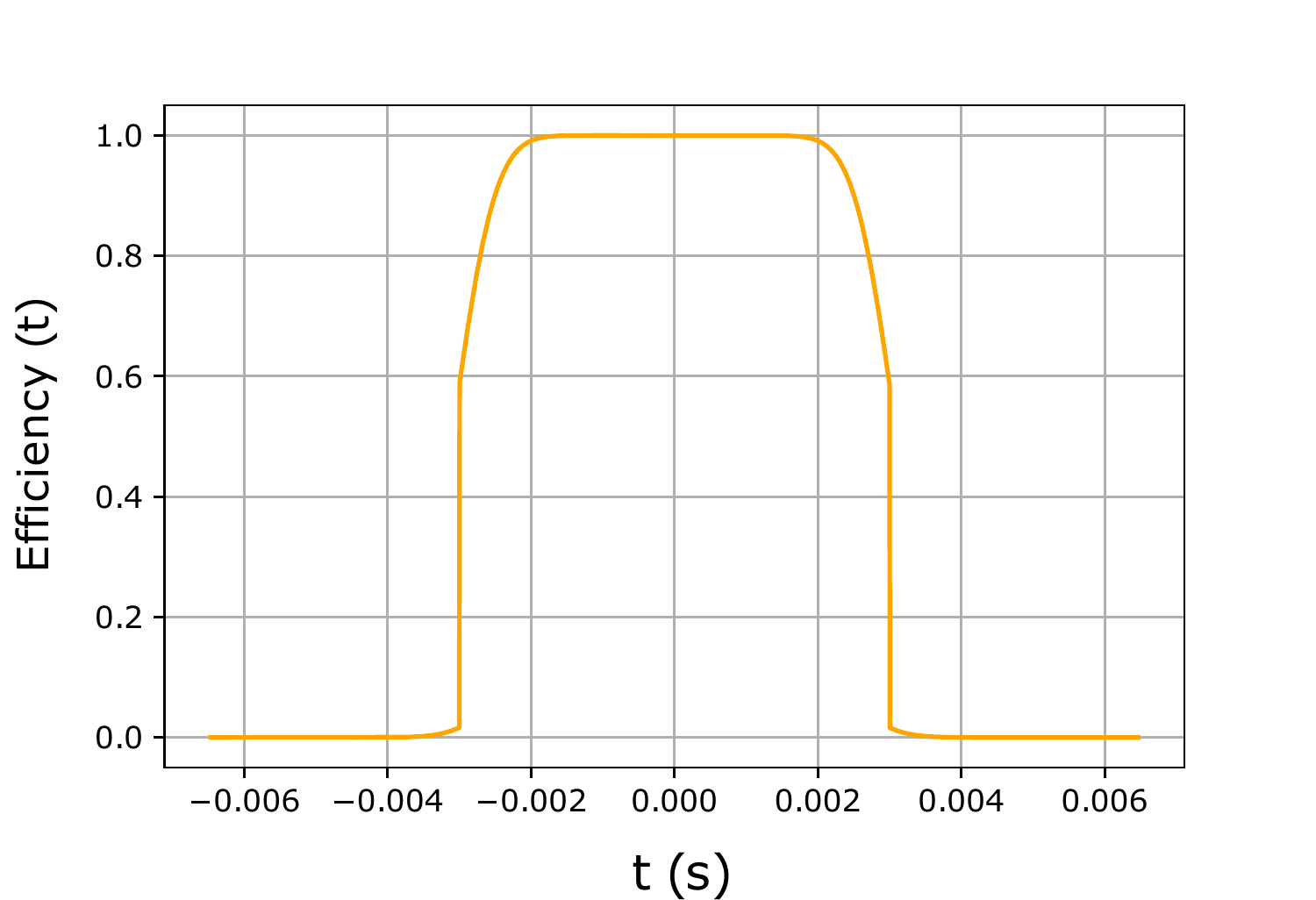}
  \caption{}
  \label{fig:sub2}
\end{subfigure}
\caption{Instantaneous channel efficiencies of Bob (a) and Eve (b) vs time for a single passage of Alice}
\label{fig:test}
\end{figure*}

For simplicity, we assume that Bob's station is positioned on the equator and Alice's and Eve's satellites have two circular orbits in the equator plane (see Fig.~\ref{fig:SatAtt}). This is an extreme simplification of Eve's task, since this guarantees that at some time Eve will be exactly in the line of sight of Alice and Bob. Indeed, if the two orbits and Bob are not in the same plane, Eve will have a hard time to intersect the beam which has a diameter below one meter close to Alice. So arguably, her best strategy would be to attach her satellites directly on Alice's. But as for the ground case, we impose an exclusion radius around Alice (in agreement with the general assumption also valid for QKD), which sets the minimal distance between Alice and Eve orbits. Note that all objects in space are well monitored and and this assumption can be verified with a reasonable effort \cite{esa_debris}.

Both Alice's and Eve's angular velocities have the same direction as Bob on earth. 
Let's consider first a single passage of the satellites over the ground station. In order to calculate $\gamma$, we trace the instantaneous collection efficiency of Bob and Eve with respect of time, where time zero is fixed such as Alice, Eve, Bob and the center of the earth are aligned. We conservatively assume that if Eve completely intersects the beam, she resends the signals to Bob and is consequently not detected directly. Finally, we integrate the instantaneous efficiency over the period of communication between Alice and Bob. The instantaneous efficiency with respect of time are calculated for Bob and Eve respectively as:

\begin{equation}
\eta^B(t) = \eta_b*\frac{D_R^2}{\theta_{div}^2d_B(t)^2}
\end{equation} 

\begin{equation}
\eta^E(t) = \frac{2}{\pi} \iint_{A(t)} \frac{e^{-2\left(\frac{x^2+y^2}{\theta_{div}^2d_E(t)^2}\right)}}{{\theta_{div}^2d_E(t)^2}} \,dx\,dy
\end{equation}

where $d_B(t)$ and $d_E(t)$ are the distances from Alice of Bob and Eve with respect of time and the integral is considered over the area of Eve telescope $A(t)$ moving with respect of Alice telescope. In the end in order to calculate $\gamma$ with evaluate the ratio between the average efficiency integrated over a period of communication between Alice and Bob:

\begin{equation}
\gamma = \frac{\int_{-T}^{T} \eta^B(t) \,dt}{\int_{-T}^{T} \eta^E(t) \,dt}
\end{equation}
where $T = 400s$ as it can be seen from Fig.\ref{fig:sub1}.

\begin{figure}[t]
\centering
\includegraphics[scale=0.55]{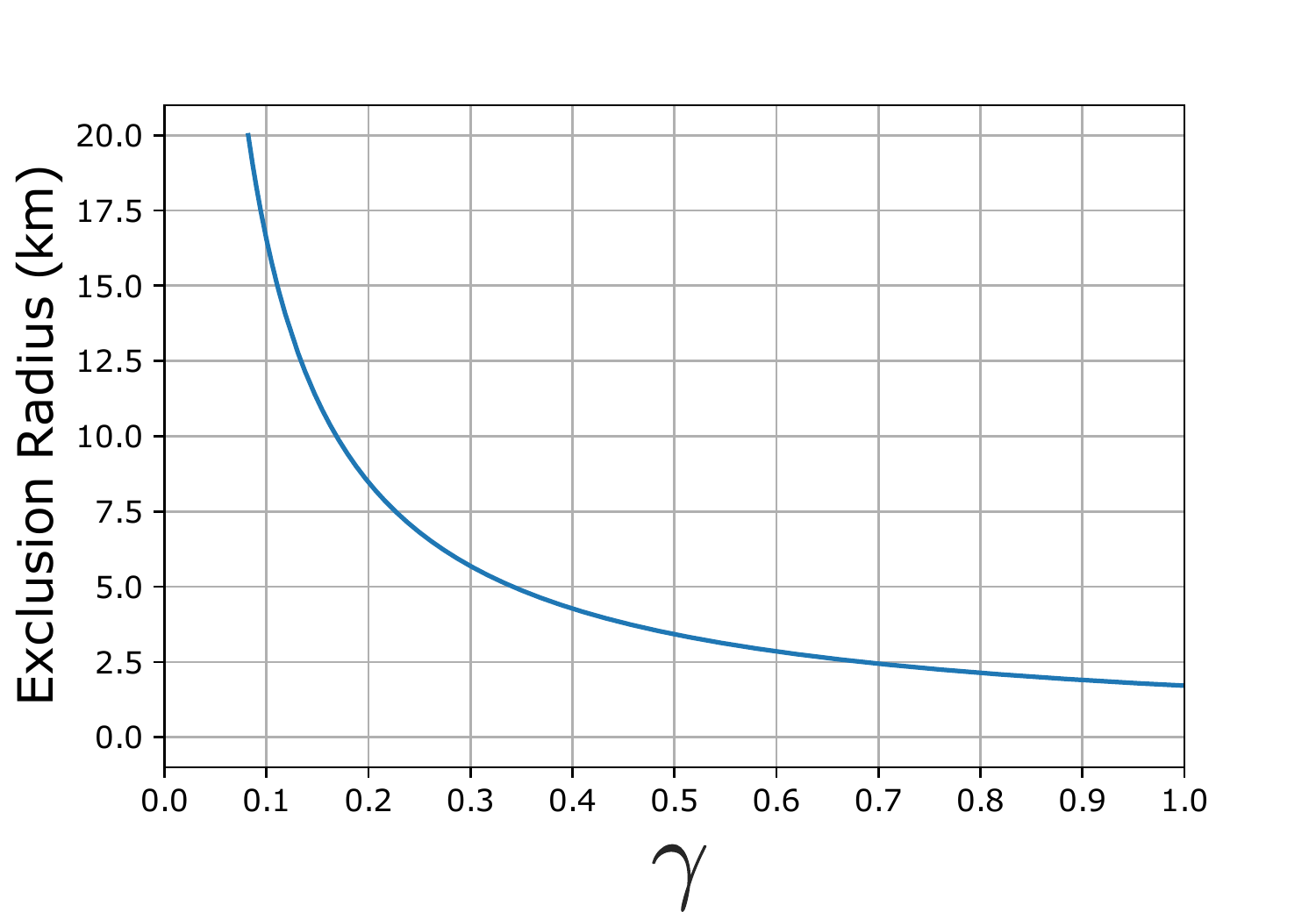}
\protect\caption{Exclusion raduis vs $\gamma$ (for a single passage).} 
\label{fig:RvsG}
\end{figure}

Fig.~\ref{fig:sub1} and Fig.~\ref{fig:sub2} show the instantaneous efficiencies of Bob and Eve, respectively, for a configuration with an exclusion radius around Alice satellite of 16 km and Alice's orbit is 600 km above ground. The communication is considered to be started and finished when the satellite is 20° above the horizon with respect to Bob. As it can be seen the communication Alice and Bob last for over 400 s, while Eve's time of interception is only of few hundreds of ms. If the efficiency is averaged over the passage of the satellite we can again calculate the factor $\gamma$. Fig.~\ref{fig:RvsG} shows the needed exclusion radius in order to obtain a given $\gamma$. We can see that for an exclusion radius above 16 km, a value of $\gamma$ lower than $0.1$ it can be assured, again. Note, this is for a size of Eve's telescope of 2m (sic!), which would be very easy to spot.  

This value of $\gamma$ is true for a single passage, however, a passage of Alice over Bob occurs at a much higher frequency than the perfect alignment of the Alice, Eve and Bob. The closer Eve is to Alice the longer the beating period will be. This means that in order to implement such an attack it is not sufficient to use a single satellite in a fixed orbit. The active strategy would involve either a constellation of satellites or a satellite with adjustable orbit. Just to give an idea, for the parameters considered here (orbit of 600 km above ground for Alice and Eve 15 km below her) the period between two links from Alice to Bob is around 1.7 hours, while the period between two intercept events for Eve (Eve is in between Alice and Bob during communication) would be of 20 days. 

Overall, these estimations, although based on a worst case scenario, show that it is possible to guarantee a $\gamma$ as low as $0.1$.

\section{\label{sec:App4}: Physical probabilistic channel model}

\begin{figure*}[tbh]
\centering
\includegraphics[scale=0.3]{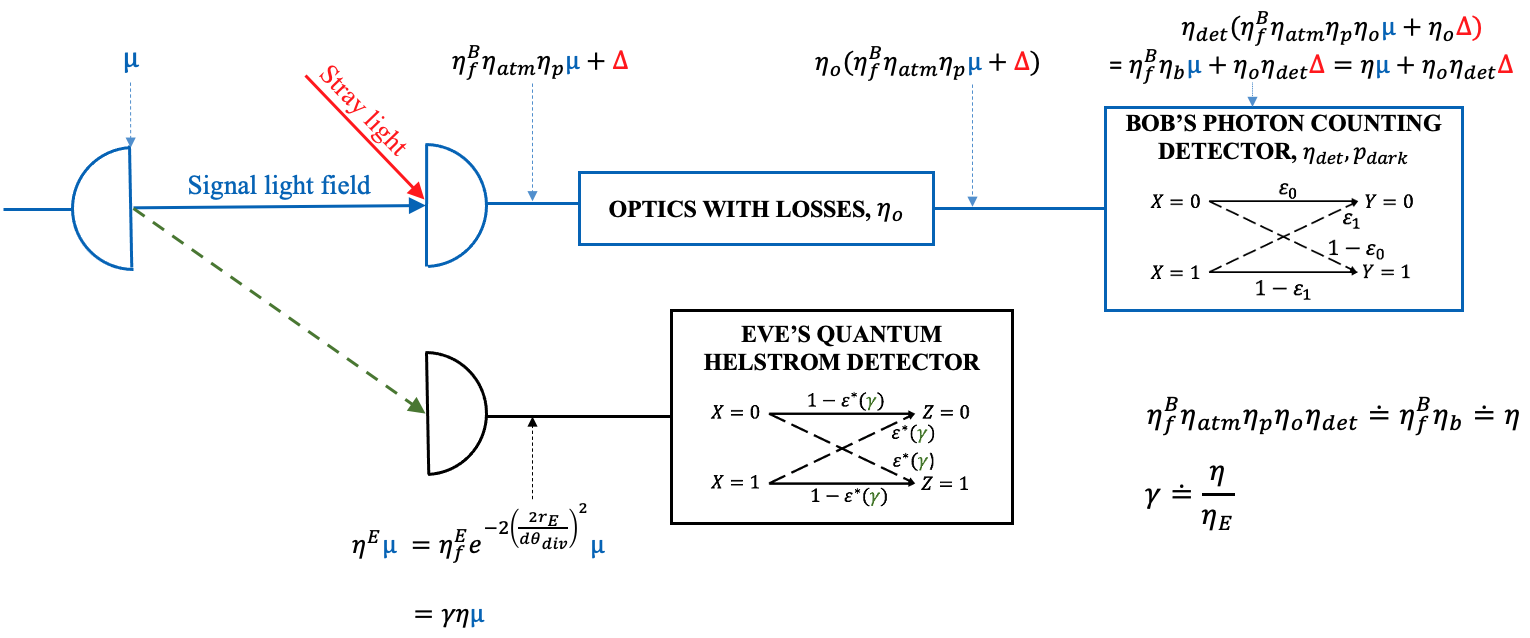}
\protect\caption{Graphical representation of our degraded quantum wiretap channel model.} 
\label{fig:Representation3}
\end{figure*}

Fig. \ref{fig:Representation3} represents our (degraded) quantum wiretap channel model, which is a physical-probabilistic model. It shows the impairments undergone by the transmitted average number of photons, $\mu=\abs{\alpha_1}^2$, over the main and wiretap channels. 

Bob's channel power attenuation has the following contributions: $\eta_f^B$ (free-space), $\eta_{atm}$ (atmospheric, e.g. turbulences), $\eta_p$ (pointing) and $\eta_o$ (optical loss from telescope input lens to detector). The detector efficiency, $\eta_{det}$, is included in the overall losses parameter $\eta$. The channel transition probabilities are also affected by the dark current noise, modelled by $p_{dark}$. Then, the probabilistic model of the detector is fully described by the probabilities $\epsilon_0=(1-p_{dark})e^{-\eta_o\Delta}$ and $\epsilon_1=(1-p_{dark})e^{-(\eta\abs{\alpha_1}^2 +\eta_o\Delta)}$.

Eve's channel is only affected by propagation in free-space, $\eta_f^E$, and by the Gaussian angular distribution of the beam, modelled as $e^{-2(\frac{2\theta_E}{\theta_{div}})^2}$, or $e^{-2(\frac{2r_E}{d\theta_{div}})^2}$, since $\theta_E \approx r_E/d$ with $\theta_E$ Eve's exclusion angle, $\theta_{div}$ the divergence angle and $d$ the distance between Alice and Bob. 
In our model, we express the number of photons at Eve's detector as a fraction of those received by Bob as $\eta^E(\gamma) = \gamma \eta$. This modelling makes the parameter $\gamma$ a design choice which controls the distinguishability of the coherent states at Eve's detector.
\section{\label{sec:App5} Distinguishability of Eve's coherent states with $\gamma$ at Eve's detection.}
\begin{figure}[tbh]
\includegraphics[scale=0.13]{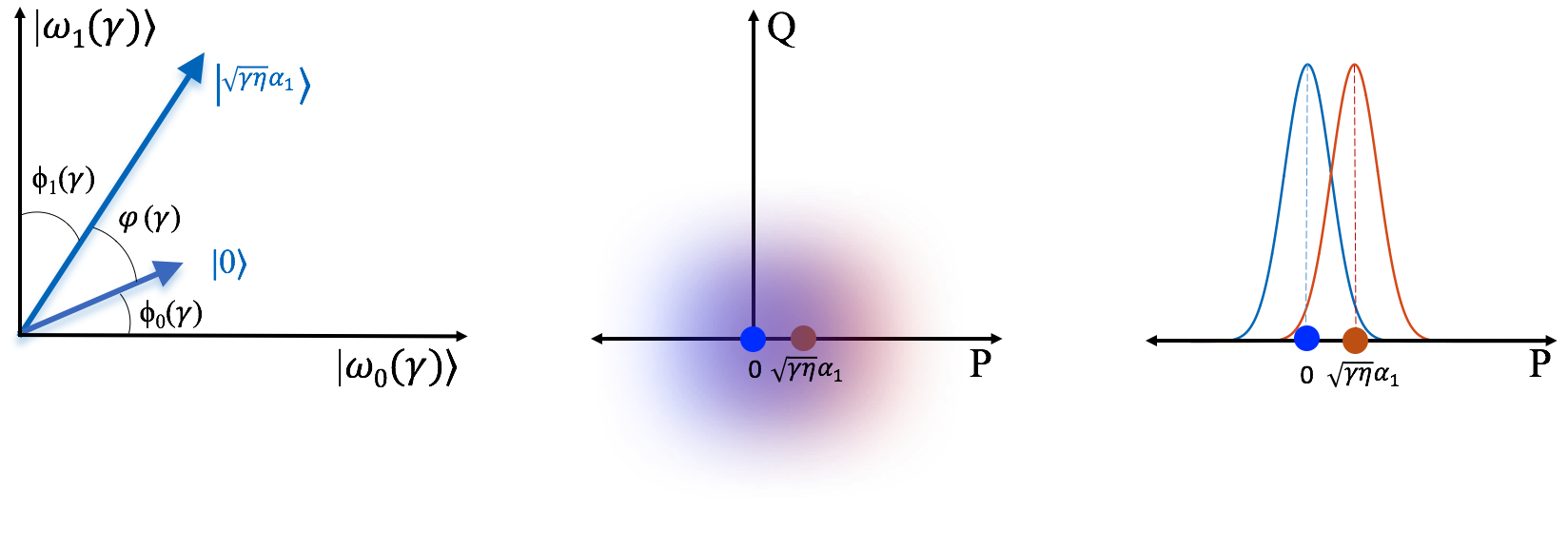}
\protect\caption{Representation of the coherent states sent by Alice and received by Eve in the qubit space, in the classical phase space and as histograms on the quadrature P (respectively from left to right). 
} 
\label{fig:DiffReps}
\end{figure}
\begin{figure}[tbh]
\includegraphics[scale=0.1]{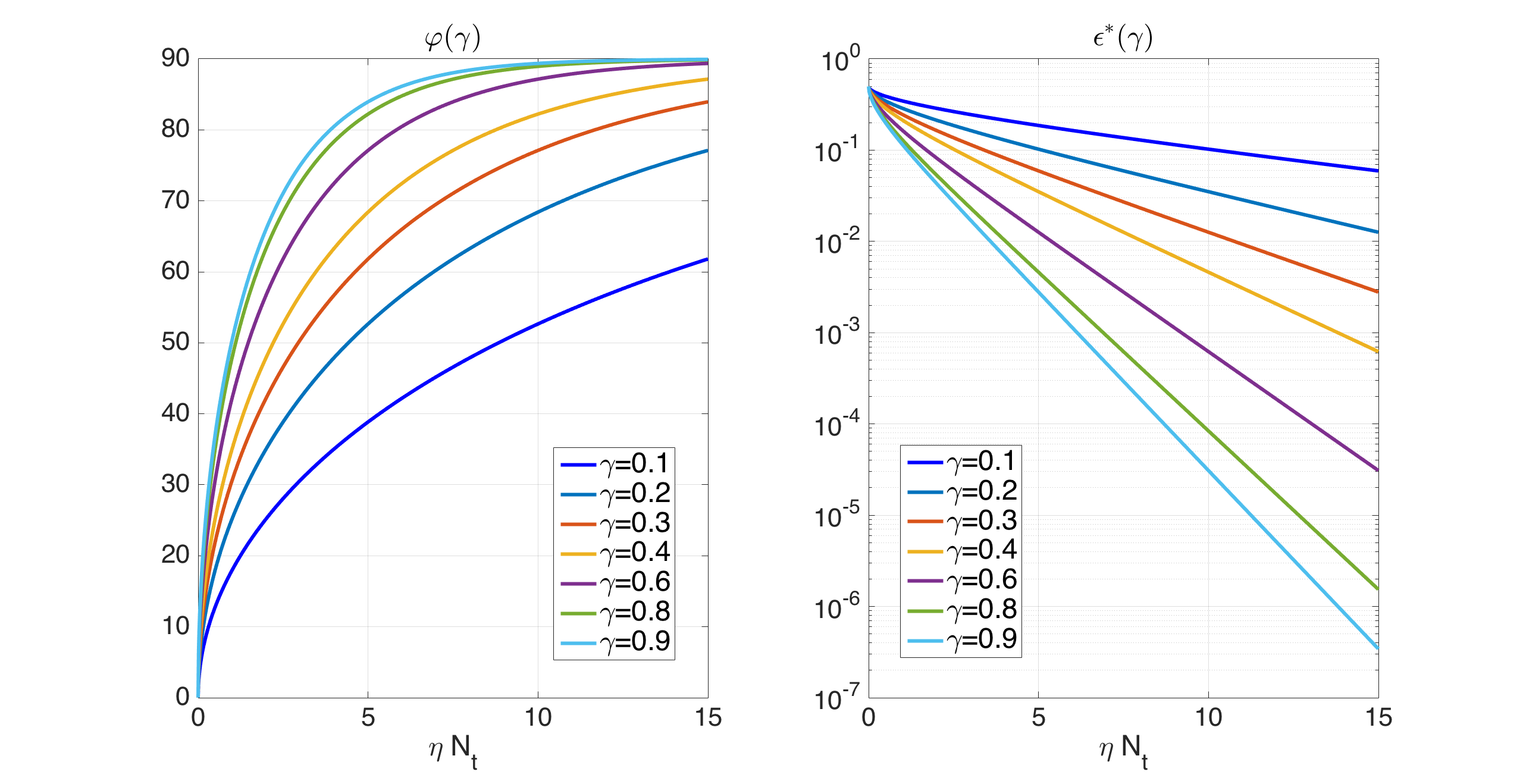}
\protect\caption{The angle $\phi(\gamma)$ (left) and the error probability $\epsilon*(\gamma)$ vs the received number of photons $\eta \mu$ for different values of $\gamma$.
} 
\label{fig:Sensitivity}
\end{figure}

Our design parameter $\gamma$ controls the distinguishability (orthogonality) of the coherent states at Eve's detection. This dependency can be visualized in different representations, as shown in Fig. \ref{fig:DiffReps}: (left) representation of the (rank-2) Hilbert space spanned by the coherent states at Eve's reception, $\varphi(\gamma)$ is the angle between the coherent states, (center) the P-Q phase space with the (unitary variance) Gaussian uncertainty of each state and (right) representation of the Gaussian distribution along the P quadrature. 
Given the angle $\varphi(\gamma)= cos^{-1}\left(\langle 0,  \sqrt{\gamma\eta} \alpha_1 \rangle \right)$, the Holevo-Helstrom optimization induces $\epsilon*(\gamma)$ and the symmetric space $\phi_0^*(\gamma)  =  \phi_1^*(\gamma) = 0.5(\pi/2 -\varphi(\gamma))$.  \\

Fig. \ref{fig:Sensitivity} shows the sensitivity of $\varphi(\gamma)$ to variations of the received number of photons. The lower the $\gamma$, the less sensitivity. We observe that for $\eta\mu = 4$, we have $\varphi(0.1)=35^o$ and $\epsilon^*(0.1)=0.2$ with an increase in orthogonality of 10\% and a decrease of 5\% in (uncoded) error probability for a fluctuation around $1$ photon.\\



\end{document}